\documentclass[letterpaper]{article} 
\usepackage{aaai25}  
\usepackage{times}  
\usepackage{helvet}  
\usepackage{courier}  
\usepackage[hyphens]{url}  
\usepackage{graphicx} 
\usepackage{natbib}  
\usepackage{caption} 
\usepackage{amsmath}
\usepackage{amssymb}
\usepackage{multirow}
\usepackage{subcaption}
\usepackage{xcolor}
\usepackage{algorithm}
\usepackage{algorithmic}

\usepackage{newfloat}
\usepackage{listings}
\DeclareCaptionStyle{ruled}{labelfont=normalfont,labelsep=colon,strut=off} 
\floatstyle{ruled}
\newfloat{listing}{tb}{lst}{}
\floatname{listing}{Listing}
\title{CAMSIC: Content-aware Masked Image Modeling Transformer \\ for Stereo Image Compression}
\author{
    Xinjie Zhang\textsuperscript{\rm 1,2}\thanks{This work was partially performed when Xinjie Zhang was an Intern at SenseTime.},
    Shenyuan Gao\textsuperscript{\rm 1}, Zening Liu\textsuperscript{\rm 1}, Jiawei Shao\textsuperscript{\rm 1,3}, \\ Xingtong Ge\textsuperscript{\rm 2}, Dailan He\textsuperscript{\rm 4}, Tongda Xu\textsuperscript{\rm 5}, Yan Wang\textsuperscript{\rm 5}, Jun Zhang\textsuperscript{\rm 1}\thanks{Corresponding Author.}
}
\affiliations{
    \textsuperscript{\rm 1}The Hong Kong University of Science and Technology \\
    \textsuperscript{\rm 2}SenseTime Research \\
    \textsuperscript{\rm 3}Institute of Artificial Intelligence (TeleAI), China Telecom \\
    \textsuperscript{\rm 4}The Chinese University of Hong Kong  \\
    \textsuperscript{\rm 5}Institute for AI Industry Research (AIR), Tsinghua University \\
    
   \{xzhangga, sygao, zhening.liu\}@connect.ust.hk, shaojw2@chinatelecom.cn, xingtong.ge@gmail.com, hedailan@link.cuhk.edu.hk, x.tongda@nyu.edu, wangyan202199@163.com, eejzhang@ust.hk 
}

\usepackage{bibentry}

\begin{document}

\maketitle

\begin{abstract}
Existing learning-based stereo image codec adopt sophisticated transformation with simple entropy models derived from single image codecs to encode latent representations. However, those entropy models struggle to effectively capture the spatial-disparity characteristics inherent in stereo images, which leads to suboptimal rate-distortion results. In this paper, we propose a stereo image compression framework, named CAMSIC. CAMSIC independently transforms each image to latent representation and employs a powerful decoder-free Transformer entropy model to capture both spatial and disparity dependencies, by introducing a novel content-aware masked image modeling (MIM) technique. Our content-aware MIM facilitates efficient bidirectional interaction between prior information and estimated tokens, which naturally obviates the need for an extra Transformer decoder. Experiments show that our stereo image codec achieves state-of-the-art rate-distortion performance on two stereo image datasets Cityscapes and InStereo2K with fast encoding and decoding speed. 
\end{abstract}

\section{Introduction}

Stereo Image Codec (SIC) compresses a pair of stereoscopic images captured from distinct viewpoints by the same camera. SIC has attracted significant interest due to the growing demand for high-quality stereo image transmission and storage across applications such as autonomous driving \cite{yin20203d}, virtual reality \cite{fehn2004depth} and video surveillance \cite{stepanov2016concept}. By leveraging the inherent correlation between views, they achieve higher coding efficiency compared with single image codecs.

\begin{figure}[t]
  \centering
  \includegraphics[scale=0.8]{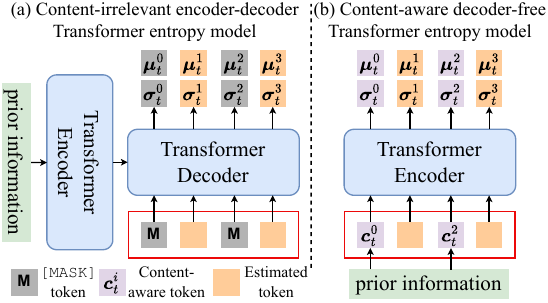}
  \caption{Comparison of two different mask image modeling Transformer entropy models. \texttt{[MASK]} token is a learnable parameter, which is irrelevant to specific image content. Tokens within the same red rectangle have bidirectional interactions. The vanilla content-irrelevant masked image modeling uses uniform \texttt{[MASK]} tokens to fill the unestimated positions and relies on an inefficient encoder-decoder Transformer to propagate the prior information. Differently, our content-aware masked image modeling replaces the uninformative \texttt{[MASK]} tokens with proposed content-aware tokens, which enables an efficient decoder-free Transformer architecture with more sufficient information interactions.}
  \label{fig:mim_comparision}
\end{figure}


Over the past decades, classical multi-view image coding standards, such as H.264-based MVC \cite{vetro2011overview} and H.265-based MV-HEVC \cite{tech2015overview}, have catalyzed the development of learning-based stereo image compression approaches \cite{liu2019dsic, deng2021deep, wodlinger2022sasic, zhai2022disparity, deng2023masic}. Given a reference view, these methods, rooted in the predictive coding paradigm, employ disparity-compensated prediction in either pixel or feature space to compress  current view. Additionally, recent advancements have introduced a bidirectional coding framework \cite{lei2022deep, wodlinger2024ecsic} that utilizes a cross-attention mechanism, further enhancing content correlation exploitation between stereo images. Albeit remarkable, the odyssey of stereo image compression research concentrates on producing compact representations through various information flows or network designs before entropy coding. However, an orthogonal direction, which reduces redundancy by devising a superior spatial-disparity entropy model for entropy coding, is rarely explored.

Previous learning-based SIC approaches often adapt existing single image compression solutions, such as hyperprior \cite{balle2018variational} and auto-regressive prior \cite{minnen2018joint}, into conditional formats. These adapted entropy models, typically based on convolutional neural network (CNN) architectures, aim to capture spatial-disparity correlations. However, CNNs are inherently limited by their local receptive fields, which hampers their ability to model long-range dependencies. This limitation often leads to a rough estimation of probability distribution and inferior compression performance, especially when the disparity distance between stereo images is large. Consequently, we argue that better coding gains can be achieved as long as we find a more effective spatial-disparity entropy model.

In this paper, we present a stereo image compression framework that centers on a powerful Transformer entropy model to leverage spatial-disparity dependencies between current and previously decoded views. This framework incorporates a straightforward image encoder-decoder pair from \citet{he2022elic} to extract latent representations for each view. As illustrated in Fig.~\ref{fig:mim_comparision} (a), we introduce the vanilla masked image modeling (MIM) style \cite{chang2022maskgit} to construct a content-irrelevant encoder-decoder Transformer entropy model. This model captures spatial information from the context of decoded tokens and disparity information from previous decoded view. Since the bi-directional self-attention in Transformer allows \texttt{[MASK]} tokens to use context from both directions, it supports a non-sequential auto-regressive decoding process and decodes all tokens in the images in a few steps, achieving a good performance-speed trade-off \cite{chang2022maskgit}.

Nevertheless, the vanilla MIM-based Transformer entropy model still exhibits two significant shortcomings. Firstly, it uses uniform \texttt{[MASK]} tokens to indicate the positions of the tokens whose probability distributions are yet to be estimated, which are irrelevant to the image content and uninformative to entropy coding. As a result, it wastes much computation and leads to inferior coding gains. Secondly, it needs an extra Transformer decoder to propagate the prior information via cross-attention. Such a complicated encoder-decoder Transformer architecture only permits an inefficient unidirectional propagation from the prior information to the input tokens of the Transformer decoder. 

To overcome these challenges, we develop a novel \textbf{C}ontent-\textbf{A}ware \textbf{M}asked image modeling Transformer entropy model for \textbf{S}tereo \textbf{I}mage \textbf{C}ompression (CAMSIC) shown in Fig.~\ref{fig:mim_comparision} (b). Specifically, we introduce content-aware tokens generated by prior information to replace the uninformative \texttt{[MASK]} tokens, forming a new content-aware masked image modeling style, which brings three major benefits. First, the interactions between the proposed content-aware tokens and the estimated tokens provide useful information about each specific position, preventing distraction of previous uninformative [MASK] tokens when updating the token features.
Second, it allows a bidirectional information flow between prior information and estimated tokens via self-attention. Compared with static prior information in the vanilla MIM, content-aware tokens, carrying specific prior information, can also be iteratively updated. Thus, it can effectively capture the context information from already estimated tokens and reduce the estimation uncertainty of remaining tokens. Third, since our design naturally embeds prior information propagation into each self-attention operation, we discard the whole Transformer decoder and devise an encoder-only Transformer architecture with low computation overhead. 
In summary, our main contributions are three-fold:
\begin{itemize}
    \item[$\bullet$] We introduce a learning-based stereo image compression framework with a simple image encoder-decoder pair, which uses an elegantly neat but powerful Transformer entropy model based on masked image modeling to exploit the relationship between the left and right images. 
    \item[$\bullet$] We present a unique content-aware masked image modeling style for Transformer entropy model. This innovation enables more effective and extensive interactions between prior information and estimated tokens, while also facilitating an efficient yet strong decoder-free Transformer entropy model architecture design. 
    \item[$\bullet$] Experimental results show that our proposed method with lower encoding and decoding latency significantly outperforms existing learning-based stereo image compression methods. Detailed ablation studies and analyses validate the effectiveness of each proposed component.
\end{itemize}

\section{Related Works}
\textbf{Single Image Compression}. 
Traditional standard image codecs, such as JPEG \cite{wallace1991jpeg}, JPEG2000 \cite{skodras2001jpeg}, BPG \cite{bpg2014}, and VVC-intra \cite{bross2021overview}, typically employ a three-step process involving transformation, quantization, and entropy coding. Regrettably, these steps are optimized independently, which may compromise overall compression performance. In contrast, recent learning-based image compression methods \cite{balle2017end, balle2018variational, minnen2018joint, cheng2020learned, chen2021end, zhu2021transformer, chen2022two, zou2022devil, he2022elic, liu2023learned} have demonstrated remarkable success, outperforming traditional methods in terms of rate-distortion (RD) performance. These modern approaches utilize nonlinear transformations (e.g., generalized divisive normalization \cite{balle2015density}, attention mechanism \cite{cheng2020learned, chen2021end}, and stacks of residual bottleneck blocks \cite{he2022elic}) to generate concise latent representations. Moreover, a series of advanced entropy models, such as factorized prior \cite{balle2017end}, hyperprior \cite{balle2018variational}, and auto-regressive context prior \cite{minnen2018joint}, are employed to accurately approximate the probability distribution of quantized latent representations. These existing works serve as foundational elements for our stereo image compression solution.


\noindent \textbf{Stereo Image Compression}. 
Traditional multi-view image codecs, such as H.264-based MVC \cite{vetro2011overview} and H.265-based MV-HEVC \cite{tech2015overview}, typically compress stereo images using disparity compensation prediction, which enjoys high compression efficiency but heavily depends on prior knowledge to design hand-crafted modules. Recently, learning-based stereo image compression methods have shown substantial improvements in compression performance. These existing works can be roughly classified into two categories: (1) Unidirectional coding \cite{liu2019dsic, deng2021deep, wodlinger2022sasic, zhai2022disparity, deng2023masic} begins with the prediction of a disparity-compensated view in either pixel or feature space, followed by the compression of residuals between the current and predicted views. (2) Bidirectional coding \cite{lei2022deep, wodlinger2024ecsic, liu2025bidirectional} incorporates a cross-attention mechanism within both the encoder and decoder to exploit the mutual information across stereo images. Apart from the above mentioned joint encoding methods, some recent works \cite{zhang2022ldmic, mital2023neural} have ventured into distributed multi-view image coding by using cross-attention alignment to achieve competitive RD performance. 
Contrary to the prevailing focus on intricate nonlinear transformations by exploring different network structures or information flows, we investigate the application of the Transformer architecture in entropy coding and propose a gracefully simple but powerful Transformer entropy model for stereo image compression. It effectively exploits spatial-disparity correlations, which eliminates the need for complex feature extraction or warping operations, thereby streamlining the compression process and establishing a neat yet potent coding framework.

\noindent\textbf{Masked Image Modeling}. 
Inspired by the masked language modeling \cite{kenton2019bert} in natural language processing, masked image modeling (MIM) has been recently applied to representation learning for vision tasks \cite{he2022masked, xie2022simmim, liu2022mixmim}. It has shown great success on various downstream tasks, including image generation \cite{chang2022maskgit, liang2022nuwa}, action recognition \cite{tong2022videomae, feichtenhofer2022masked}, video compression \cite{xiang2023mimt} and video prediction \cite{gupta2022maskvit}. In this paper, we apply the MIM technique to stereo image compression. Note that the vanilla MIM style uses a uniform \texttt{[MASK]} token to take the places of the tokens whose probability distribution are not yet estimated, which wastes much computation on the content-irrelevant \texttt{[MASK]} tokens, thus leading to sub-optimal coding gains. In addition, it resorts to an extra Transformer decoder to propagate the prior information in a unidirectional manner. To address these limitations, we develop a novel content-aware MIM style, which replaces the uninformative \texttt{[MASK]} tokens with content-aware tokens. It not only enables a bidirectional interaction between the prior information and estimated tokens, but also allows us to design an efficient decoder-free Transformer entropy model. 

\noindent \textbf{Transformer Architecture}. Recent advancements in Transformer architectures have primarily employed two configurations: (i) The encoder-decoder Transformer is tailored for sequence-to-sequence tasks such as machine translation and question answering. Despite its flexibility in generating varying output lengths, it is computationally intensive and requires substantial resources for training due to its dual-module structure. (ii) The decoder-free Transformer is suitable for input understanding tasks (e.g., sentence matching and feature extraction). This structure usually offers quicker training and lower computational demands.
Given the effectiveness of Transformer in information fusion and interaction, recent compression practices like VCT \cite{mentzervct} and MIMT \cite{xiang2023mimt} have incorporated encoder-decoder Transformer architectures into entropy models, significantly advancing compression performance. By contrast, we leverage a decoder-free Transformer to achieve a better speed-performance trade-off.

\begin{figure}[t]
  \centering
  \includegraphics[scale=0.72]{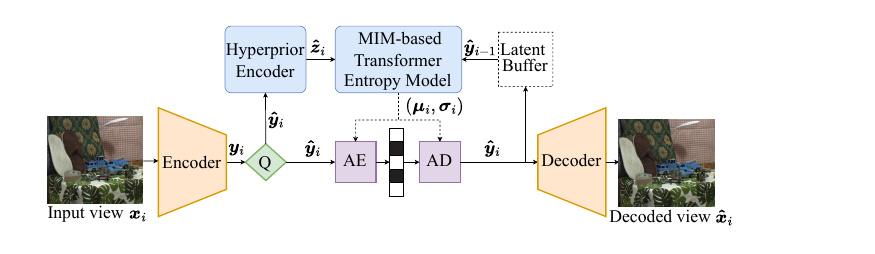}
  \caption{Our overall stereo image compression framework. Given an input image $\boldsymbol{x}_v$, we first encode the input view to produce the latent representation $\boldsymbol{y}_v$ that is quantized to $\boldsymbol{\hat{y}}_v$. To transmit $\boldsymbol{\hat{y}}_v$ with fewer bits, an MIM-based Transformer entropy model is introduced to predict a probability distribution for $\boldsymbol{\hat{y}}_v$ given hyperprior $\boldsymbol{\hat{z}}_v$, and previously stored representations $\boldsymbol{\hat{y}}_{v-1}$. Finally, the quantized representation $\boldsymbol{\hat{y}}_v$ is decoded to the reconstructed view $\boldsymbol{\hat{x}}_v$. When compressing the first view, a learned embedding $\boldsymbol{\hat{y}}_{-1}$ is used. The arithmetic encoder and decoder after the hyperprior encoder are omitted for brevity.} 
  \label{fig:ec_arch}
\end{figure}

\begin{figure*}[t]
  \centering
  \includegraphics[scale=0.74]{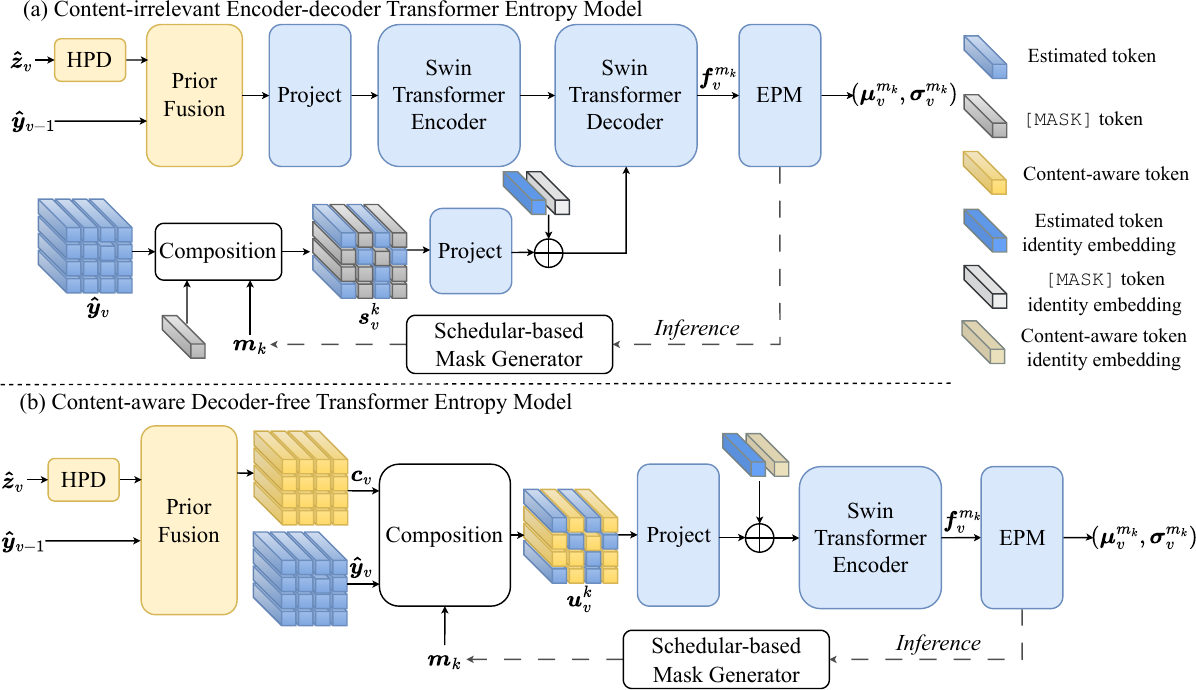}
  \caption{Structures of content-irrelevant encoder-decoder Transformer entropy model and our content-aware decoder-free Transformer entropy model. HPD indicates hyperprior decoder. EPM means entropy parameter module. The mask $\boldsymbol{m}_k$ starts at zero and is then updated iteratively by a scheduler-based generator during inference. In each iteration, a sinusoidal scheduler computes the number of tokens $n_k$ to be encoded/decoded. The mask generator then sets the values of the $n_k$ tokens with the lowest estimated bitrates and the previously encoded/decoded tokens as 1 to update $\boldsymbol{m}_k$. (a) By using $\boldsymbol{m}_k$ to mix the \texttt{[MASK]} token and $\boldsymbol{\hat{y}}_v$, a composite context $\boldsymbol{s}_v^k$ is synthesized and sent to a projection module, where two identity embeddings are added to further different token types. Both the prior information and the composite context are input into a Transformer decoder to estimate the probability distribution parameters $(\boldsymbol{\mu}_v^{m_k}, \boldsymbol{\sigma}_v^{m_k})$ for the masked tokens. (b) The \texttt{[MASK]} token is replaced by the context-aware tokens $\boldsymbol{c}_v$ to form the composite context $\boldsymbol{u}_v^k$. This context is then input into a projection module with two identity embeddings and a Transformer encoder to output the estimated Gaussian distribution parameters $(\boldsymbol{\mu}_v^{m_k}, \boldsymbol{\sigma}_v^{m_k})$.}
  \label{fig:entropy_model}
\end{figure*}

\section{Method}

\subsection{Overview of CAMSIC}\label{sec:overview}
Our stereo image compression framework CAMSIC is shown in Fig.~\ref{fig:ec_arch}. Let $\mathcal{X}=\{\boldsymbol{x}_1, \boldsymbol{x}_2\}$ denotes a pair of stereo images, where $\boldsymbol{x}_v \in \mathbb{R}^{H \times W \times 3}$ is the image at view $v$ ($v=1,2$) with height $H$ and width $W$. 
An image encoder $E$ is used to compress each individual view $\boldsymbol{x}_v$ to a latent representation $\boldsymbol{y}_v \in \mathbb{R}^{h \times w \times d}$, where $h=H/16$, $w=W/16$ and $d$ is the latent dimension. 
$\boldsymbol{y}_v$ is then quantized to $\boldsymbol{\hat{y}}_v$ by the quantizer $Q$. Finally, we feed $\boldsymbol{\hat{y}}_v$ to an image decoder $D$ to reconstruct the view $\boldsymbol{\hat{x}}_v$. Formally, the overall coding procedure can be formulated as:
\begin{equation}
\begin{aligned}
\label{equation:compression_procedure}
    \boldsymbol{y}_v = E(\boldsymbol{x}_v; \boldsymbol{\phi}), \boldsymbol{\hat{y}}_v = Q(\boldsymbol{y}_v), \boldsymbol{\hat{x}}_v = D(\boldsymbol{\hat{y}}_v; \boldsymbol{\theta})
\end{aligned}
\end{equation}
where $\boldsymbol{\phi}$ and $\boldsymbol{\theta}$ are learnable parameters of the image encoder and decoder, respectively. To further reduce the bits to be transmitted, the quantized representation $\boldsymbol{\hat{y}}_v$ is losslessly compressed by entropy coding, where the probability distribution of $\boldsymbol{\hat{y}}_v$ is estimated by a conditional entropy model. Given the spatial context $\boldsymbol{s}_v$, hyperprior $\boldsymbol{\hat{z}}_v$, and disparity prior $\boldsymbol{\hat{y}}_{v-1}$, we model the probability of quantized representation $\boldsymbol{\hat{y}}_v$ as a Gaussian distribution:
\begin{equation}
\begin{aligned}
p_{\boldsymbol{\hat{y}}_v}(\boldsymbol{\hat{y}}_v|\boldsymbol{\hat{y}}_{v-1}, \boldsymbol{\hat{z}}_v, \boldsymbol{s}_v) = \prod_{i} \big(\mathcal{N}(\boldsymbol{\mu}_{v}, \boldsymbol{\sigma}_{v}^{2})*\mathcal{U}(-\frac{1}{2}, \frac{1}{2})\big)(\hat{y}_v^i)
\end{aligned}
\end{equation}
where $\mathcal{U}(\cdot)$ is uniform noise, $*$ denotes convolution,  $\boldsymbol{\mu}_{v}$ and $\boldsymbol{\sigma}_{v}^{2}$ are the estimated means and variances of the Gaussian distributions.
When encoding the first view, we expand a learned embedding with the dimension as $\mathbb{R}^{1 \times 1 \times d}$ to get $\boldsymbol{\hat{y}}_{-1} \in \mathbb{R}^{h \times w \times d}$. In this paper, we refer to prior information as the combination of disparity prior $\boldsymbol{\hat{y}}_{v-1}$ and hyperprior $\boldsymbol{\hat{z}}_{v}$ to distinguish the spatial context from other priors. By leveraging complementary priors, the conditional entropy model can estimate more accurate probability distributions.

\subsection{Content-irrelevant Encoder-decoder Transformer Entropy Model}\label{sec:mim entropy model}
\noindent \textbf{Content-irrelevant Masked Image Modeling}. The input of the Transformer is a sequence of tokens. With some abuse of notation, we still refer to these tokens as $\boldsymbol{\hat{y}}_v$, which is flattened from the latent representation. Given the quantized latent tokens $\boldsymbol{\hat{y}}_v$, the content-irrelevant MIM defines the spatial context $\boldsymbol{s}_v$ as
\begin{equation}
\begin{aligned}
    \boldsymbol{s}_v=\boldsymbol{\hat{y}}_v \odot \boldsymbol{m} + {\texttt{[MASK]}} \odot (1-\boldsymbol{m})
\end{aligned}
\end{equation}
where $\odot$ denotes element-wise multiplication and $\boldsymbol{m}$ is a Boolean mask whose value is assigned as 
\begin{equation}
\begin{aligned}
    m_i=\begin{cases} 1 & \text{if position $i$ is already estimated} \\ 0 & \text{if position $i$ is to be estimated} \end{cases}
\end{aligned}
\end{equation}
Based on this assignment, \texttt{[MASK]} tokens take up the positions of the tokens whose probability distributions are to be estimated. 

\noindent \textbf{Encoder-decoder Transformer}. Since these \texttt{[MASK]} tokens are irrelevant to the image content, a Transformer decoder is required to propagate the content-related prior information to the \texttt{[MASK]} tokens through cross-attention. Fig.~\ref{fig:entropy_model} (a) shows the architecture of an encoder-decoder Transformer entropy model based on content-irrelevant MIM. Specifically, we use a prior fusion network and a Transformer encoder to integrate the hyperprior $\boldsymbol{\hat{z}}_v$ and disparity prior $\boldsymbol{\hat{y}}_{v-1}$ for producing prior features. During training, we randomly generate a Boolean mask to composite the spatial context $\boldsymbol{s}_v$. The prior features and unmasked spatial tokens convey their information to the \texttt{[MASK]} tokens via cross-attention and self-attention in the Transformer decoder, respectively. Finally, we feed the feature $\boldsymbol{f}_v^{m}$ produced by the Transformer decoder into an entropy parameter module to estimate the probability distribution parameters $(\boldsymbol{\mu}_v^m, \boldsymbol{\sigma}_v^m)$.

\begin{figure}[t]
  \centering
  \subfloat
  {\includegraphics[scale=0.201]{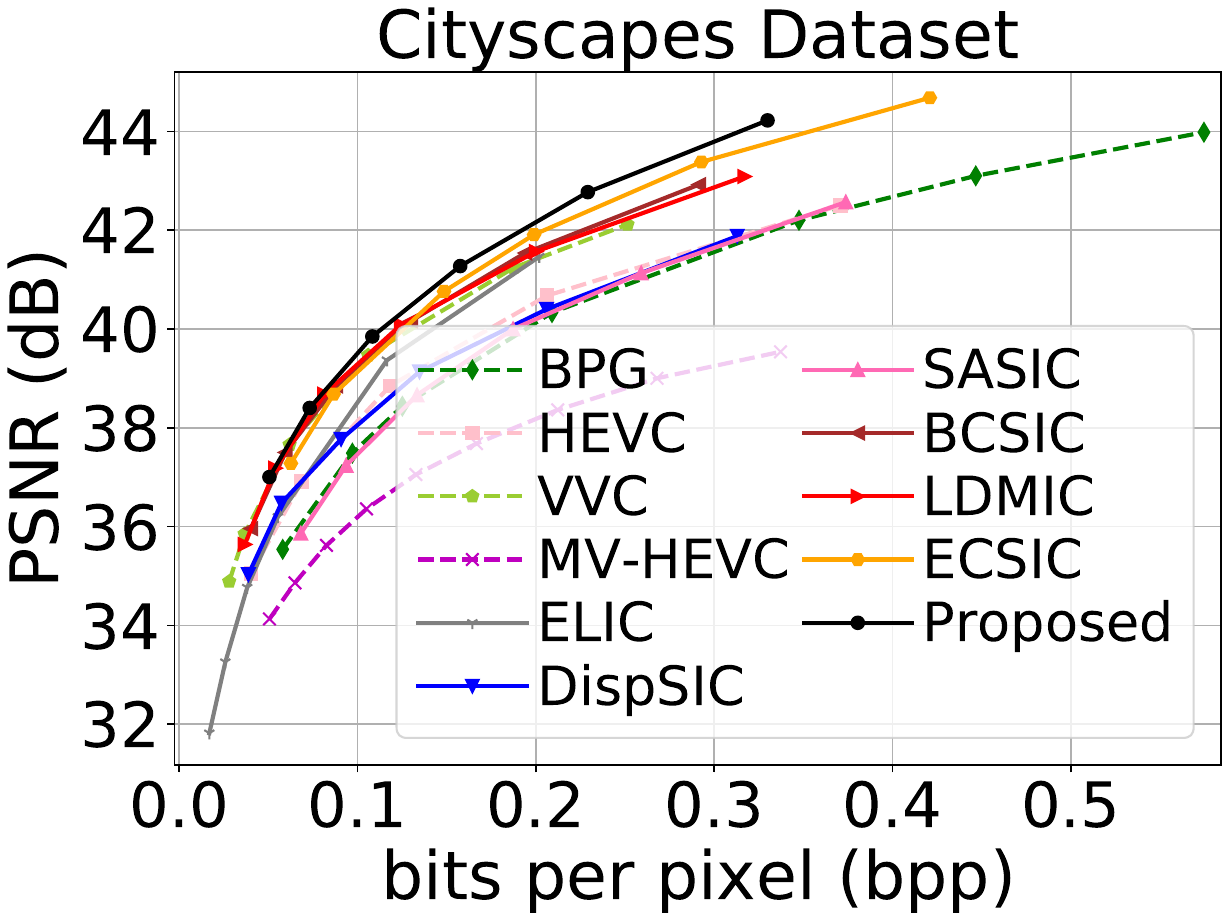}}
  \subfloat
  {\includegraphics[scale=0.201]{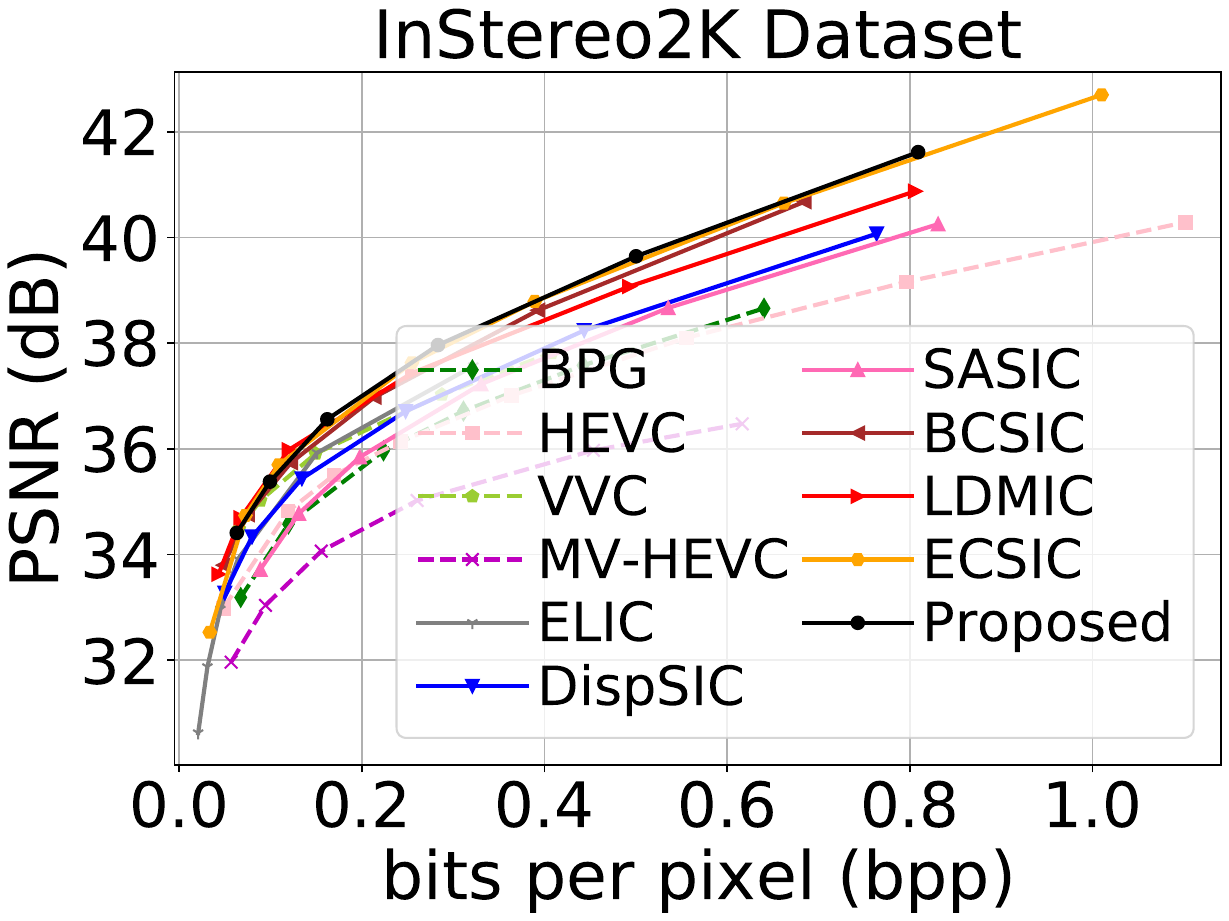}} \\
  \subfloat
  {\includegraphics[scale=0.201]{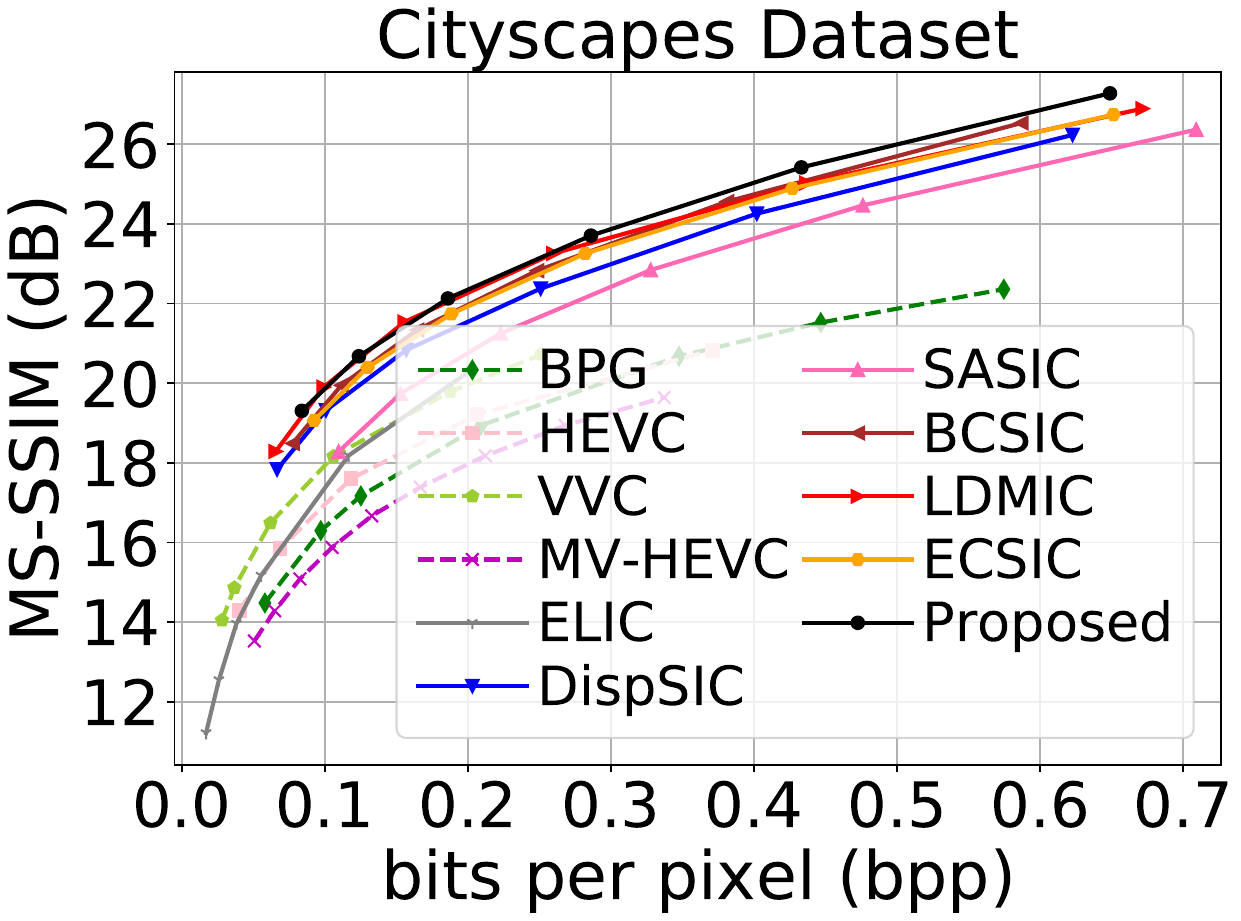}}
  \subfloat
  {\includegraphics[scale=0.201]{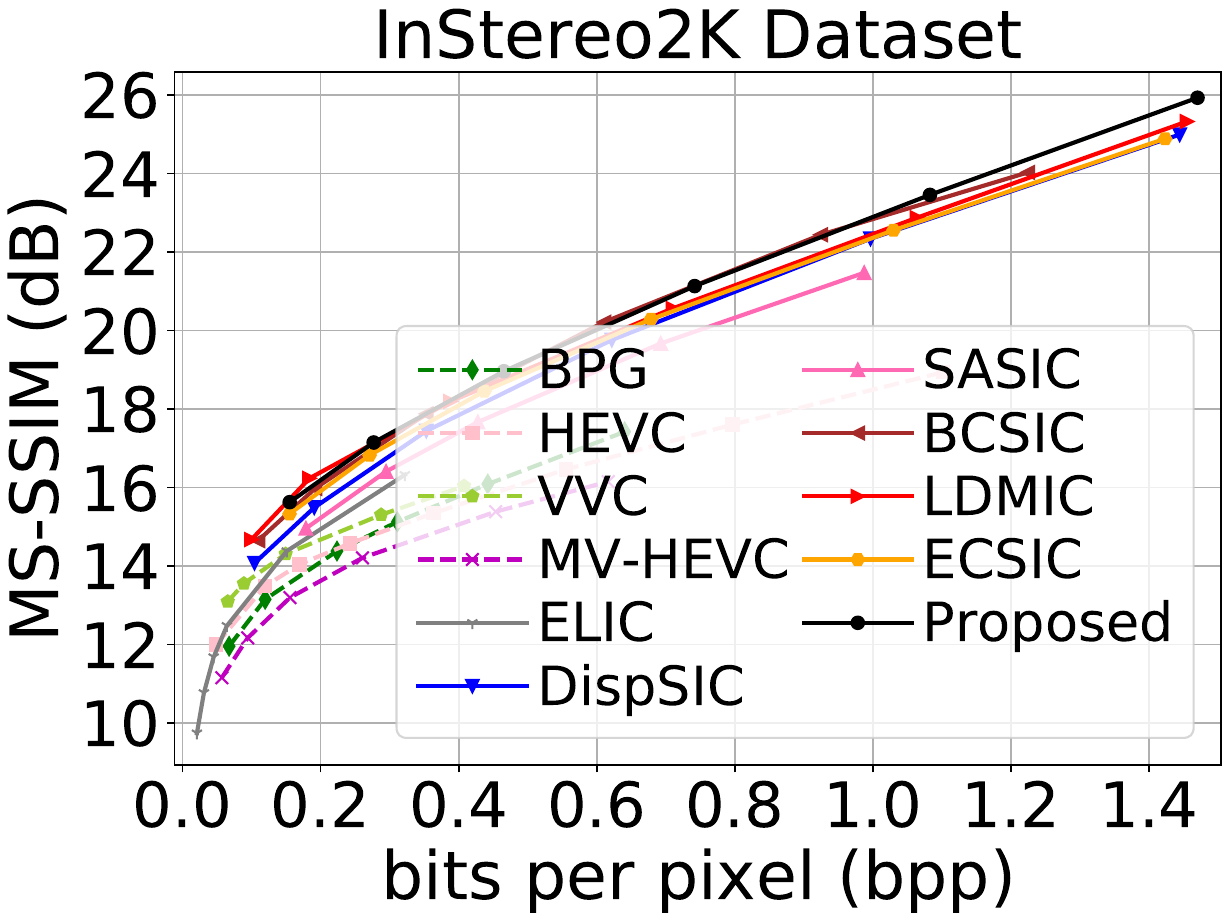}}
  \caption{Rate-distortion curves of our approach and different baselines on Cityscapes and InStereo2K datasets.}
  \label{fig:rd_performance}
\end{figure}

\subsection{Content-aware Decoder-free Transformer Entropy Model}\label{sec:content-aware fusion}
\noindent \textbf{Content-aware Masked Image Modeling}. The vanilla MIM uses a content-irrelevant \texttt{[MASK]} token to occupy the places of the tokens whose probability distribution are not yet estimated, which is not the most efficient and effective choice for entropy coding. On the one hand, interacting with the uninformative \texttt{[MASK]} tokens may distract the specific features of the estimated tokens during the repetitive self-attention operations. On the other hand, the encoder-decoder Transformer architecture is complicated but only allows a unidirectional propagation from prior information to the input tokens of the Transformer decoder. The prior information is fixed once extracted and cannot utilize the updated spatial context information from the already estimated tokens. Consequently, it is difficult for the entropy model to make full use of the power of Transformer architecture.

To overcome these limitations, we propose content-aware tokens to unleash the potential of the Transformer entropy model. Specifically, the pre-acquired prior information, including the disparity prior $\boldsymbol{\hat{y}}_{v-1}$ and hyperprior $\boldsymbol{\hat{z}}_v$, is used to generate the content-aware tokens $\boldsymbol{c}_v$. Then, a composite context $\boldsymbol{u}_v$ is created by filling the unestimated positions with content-aware tokens:
\begin{equation}
\begin{aligned}
    \boldsymbol{u}_v=\boldsymbol{\hat{y}}_v \odot \boldsymbol{m} + \boldsymbol{c}_v \odot (1-\boldsymbol{m})
\end{aligned}
\end{equation}
This novel content-aware MIM style brings multiple advantages. First, since our proposed content-aware tokens in place have already carried specific prior information about each position, they can provide a more informative context to update the features of the estimated tokens. Second, performing self-attention over the composite context can bridge a bidirectional information flow between the prior information and estimated tokens. It allows the prior information to be iteratively updated by the features of the previously estimated token features. Thus, the composite context becomes more and more concrete as the number of the estimated tokens gradually increases, thereby effectively reducing the estimation uncertainty of the remaining tokens. Third, as the prior information propagation is seamlessly embedded into each self-attention operation, it facilitates our design of a new decoder-free Transformer entropy model.

\noindent \textbf{Decoder-free Transformer.} Since the prior information has sufficiently interacted with the estimated tokens in the Transformer encoder, an explicit Transformer decoder is no longer needed. Thus, we discard the whole Transformer decoder and build an encoder-only Transformer architecture. As shown in Fig.~\ref{fig:entropy_model} (b), it comprises two parts, i.e., a prior generation network that produces content-aware tokens $\boldsymbol{c}_v$ based on different priors and a decoder-free Transformer that estimates the probability distributions of tokens. Note that we enable relative position embedding \cite{liu2021swin} in the Transformer and assign two identity embeddings to the token sequence to further distinguish different types of tokens, which effectively promote the token interactions.

\subsection{Training and Inference} \label{sec:training_and_inference}
\textbf{Bitrate Estimation.} The training phase involves a two-step prediction procedure to estimate the probability distributions of all tokens. Taking the content-aware MIM method as an example, a Boolean mask $\boldsymbol{m}$ is randomly generated to simulate various possible masking scenarios. Initially, purely based on the content-aware tokens $\boldsymbol{c}_v$, we estimate the bitrate $R_{\boldsymbol{c}_v}(\boldsymbol{\hat{y}}_v)$ of the initial tokens where the masked values are $1$. Subsequently, a composite context $\boldsymbol{u}_v$ is created to estimate the bitrate $R_{\boldsymbol{u}_v}(\boldsymbol{\hat{y}}_v)$ of the remaining tokens. Hence, the bitrate estimation based on the content-aware MIM is expressed as:
\begin{equation}
\begin{aligned}
\label{equation:PMF}
    R(\boldsymbol{\hat{y}}_v)&= R_{\boldsymbol{c}_v}(\boldsymbol{\hat{y}}_v) + R_{\boldsymbol{u}_v}(\boldsymbol{\hat{y}}_v) \\
    &=\mathbb{E}[-\log_2 p_{\boldsymbol{\hat{y}}_v}(\boldsymbol{\hat{y}}_v|\boldsymbol{c}_{v})]+\mathbb{E}[-\log_2 p_{\boldsymbol{\hat{y}}_v}(\boldsymbol{\hat{y}}_v|\boldsymbol{u}_v)]
\end{aligned}
\end{equation}
Similarly, for the content-irrelevant MIM, we replace the above content-aware token with a \texttt{[MASK]} token to estimate the bitrate during training.

During inference, the probability distributions of the tokens are iteratively estimated following a sinusoidal decoded ratio scheduler \cite{chang2022maskgit}. Let $R(\hat{\boldsymbol{y}}_{v}^{i,k})$ denotes the estimated bitrate of $i$-th token in $\hat{\boldsymbol{y}}_{v}$ during $k$-th iteration. As shown in Fig.~\ref{fig:entropy_model}, the mask $\boldsymbol{m}_0$ is initially zero during inference. In each iteration, we use the composite context $\boldsymbol{u}_v^k$ to estimate the bitrate $R(\hat{\boldsymbol{y}}_{v}^{i,k})$. The mask generator employs a scheduling function to calculate the number of tokens $n_k$ to be encoded/decoded, and selects those $n_k$ tokens that consume the fewest bits to update the mask $\boldsymbol{m}_{k+1}$, setting the values in the $n_k$ tokens and previous encoded/decoded tokens to 1. Finally, $\boldsymbol{u}_v^{k+1}$ integrates $\hat{\boldsymbol{y}}_v$ and $\boldsymbol{c}_v$ using $\boldsymbol{m}_{k+1}$ for next iteration. This iterative process ensures consistency between encoder and decoder, producing identical $(\boldsymbol{\mu}^{m_k}_v, \boldsymbol{\sigma}^{m_k}_v)$. More details about iterative decoding process are available in the appendix. 


\noindent \textbf{Quantization.} During training, since the quantization operation is not differentiable, we leverage a mixed quantizer to allow end-to-end optimization. To be specific, it adds $\boldsymbol{y}_v$ and a uniform noise $\mathcal{U}(-\frac{1}{2}, \frac{1}{2})$ together as the input to the entropy model, but rounds $\boldsymbol{y}_v$ as the input to the decoder based on straight-through estimation \cite{bengio2013estimating}. We refer readers to \citet{minnen2020channel} for more details.

\noindent \textbf{Loss Function.} The objective of our proposed method is to optimize the rate-distortion trade-off. Thus, a training loss made up of two metrics is utilized:
\begin{equation}
  \begin{aligned}
    \mathcal{L} = \lambda \sum_{v=1}^{2} d(\boldsymbol{x}_v, \boldsymbol{\hat{x}}_v) + \sum_{v=1}^{2} \big( R(\boldsymbol{\hat{y}}_v) + R(\boldsymbol{\hat{z}}_{v}) \big) \label{eq:loss_func}
  \end{aligned}
\end{equation}
where $d(\boldsymbol{x}_v, \boldsymbol{\hat{x}}_v)$ represents the distortion between the original image $\boldsymbol{x}_v$ and reconstructed image $\boldsymbol{\hat{x}}_v$ under a given metric such as MSE or MS-SSIM \cite{wang2003multiscale}. $R(\boldsymbol{\hat{y}}_v)$ denotes the estimated bitrate of the quantized representation $\boldsymbol{\hat{y}}_v$ and $R(\boldsymbol{\hat{z}}_{v})$ denotes the estimated bitrate of the hyper latent $\boldsymbol{\hat{z}}_{v}$. $\lambda$ is a hyper parameter that is used to balance the compression rate and distortion.

\section{Experiments}
\subsection{Experimental Setup}
\noindent \textbf{Dataset.} We evaluate the coding efficiency of our proposed CAMSIC on two public stereo image datasets: 
(1) Cityscapes \cite{cordts2016cityscapes}: A dataset of urban outdoor scenes with distant views. It includes 5000 image pairs at a resolution of 2048$\times$1024 pixels. We divide these into 2975 pairs for training, 500 pairs for validation, and the remaining 1525 pairs for testing.
(2) InStereo2K \cite{bao2020instereo2k}: A dataset of indoor scenes with close views. It contains 2060 image pairs at a resolution of 1080$\times$860 pixels. We allocate 2010 and 50 stereo image pairs for training and testing, respectively.
We randomly crop the images into $256\times256$ patches during training, and evaluate different codecs on full-resolution images. When the resolution of the source image is not supported, we apply the replicate padding to the right and bottom sides of the image. 


\noindent \textbf{Evaluation Metrics.} We use bit per pixel (bpp) to measure the number of bits for latent representation compression. Two popular image quality assessment metrics, namely, PSNR and MS-SSIM \cite{wang2003multiscale}, are used to evaluate the distortion between the reconstructed and original images. We also report the Bjontegaard Delta bitrate (BD-rate) \cite{2001Calculation} results to show average bitrate savings at equivalent distortion levels. In addition, both BD-PSNR and BD-MSSSIM are utilized to indicate average image quality improvements at constant bitrate.

 \begin{table}[t]
  \fontsize{7.2}{9}\selectfont
    \centering
    \begin{subtable}[t]{1\linewidth}
    \centering
        \begin{tabular}{c|cc|cc}
            \hline
            Methods & BD-PSNR $\uparrow$ & BD-rate $\downarrow$ & BD-MSSSIM $\uparrow$ & BD-rate $\downarrow$   \\
            \hline
            HEVC  & 0.499dB & -13.533\% & 0.465dB & -14.453\% \\
            VVC  & 1.621dB & {\color{blue}\underline{-37.213\%}} & 1.538dB &  -38.075\% \\
            MV-HEVC  & -1.693dB &  61.915\% & -0.736dB & 24.305\% \\
            ELIC  & 1.128dB & -25.196\% & 1.198dB & -28.346\%  \\
            DispSIC  & 0.402dB & 11.865\% & 2.948dB & -56.670\%  \\
            SASIC  & -0.021dB & 0.661\% & 2.208dB & -43.415\% \\
            BCSIC  & 1.605dB & -35.891\% & 3.355dB &  -60.292\%  \\
            LDMIC & 1.600dB & -36.264\% &  {\color{blue}\underline{3.573dB}} & {\color{blue}\underline{-64.398\%}} \\
            ECSIC  & {\color{blue}\underline{1.700dB}} & -36.665\% & 3.274dB & -59.543\%  \\
            Proposed & {\color{purple}\textbf{2.019dB}} & {\color{purple}\textbf{-41.961\%}} & {\color{purple}\textbf{3.715dB}} & {\color{purple}\textbf{-64.519\%}}\\
            \hline
        \end{tabular}
    \caption{Cityscapes dataset}
    \end{subtable}\\
    \begin{subtable}[t]{1\linewidth}
    \centering
        \begin{tabular}{c|cc|cc}
            \hline
            Methods & BD-PSNR $\uparrow$ & BD-rate $\downarrow$ & BD-MSSSIM $\uparrow$ & BD-rate $\downarrow$   \\
            \hline
            HEVC  & 0.103dB & -4.784\% &  0.092dB &  -4.260\% \\
            VVC &  0.802dB & -29.125\% & 0.669dB & -26.483\% \\
            MV-HEVC  & -1.274dB & 77.154\% & -0.578dB & 28.493\% \\
            ELIC  & 0.823dB &  -29.748\% & 0.807dB & -30.453\%  \\
            DispSIC  & 0.617dB & -23.084\% & 1.740dB & -46.916\%  \\
            SASIC  & 0.231dB & -7.825\% & 1.499dB &  -39.693\% \\
            BCSIC  & 1.249dB & -39.933\% & 2.228dB & -55.022\%  \\
            LDMIC &  1.282dB & -43.522\% & {\color{blue}\underline{2.333dB}} & {\color{purple}\textbf{-59.167\%}} \\
            ECSIC  & {\color{blue}\underline{1.428dB}} & {\color{blue}\underline{-43.551\%}} & 2.149dB & -52.587\%  \\
            Proposed & {\color{purple}\textbf{1.440dB}} & {\color{purple}\textbf{-43.608\%}} & {\color{purple}\textbf{2.426dB}} & {\color{blue}\underline{-56.773\%}} \\
            \hline
        \end{tabular}
    \caption{InStereo2K dataset}
    \end{subtable}
    \label{tab:array}
    \caption{Comparison of various codecs in BD-PSNR, BD-MSSSIM, BD-rate results relative to BPG, with the best results in {\color{purple}\textbf{bold}} and second-best ones in {\color{blue}\underline{underlined}}.}
    \label{table:rd_performance}
\end{table}

\noindent \textbf{Implementation Details.} Leveraging CompressAI \cite{begaint2020compressai}, we train our models with 6 different $\lambda$ values (256, 512, 1024, 2048, 4096, 8192 for the MSE metric; 8, 16, 32, 64, 128, 256 for the MS-SSIM metric). For MSE-optimized models, they are trained for 400 epochs with the Adam optimizer \cite{kingma2015adam}. The batch size is set as $4$. The initial learning rate is $1e^{-4}$ and decayed by a factor of $2$ every $100$ epochs.  To accelerate the experiments, both the image encoder and decoder utilize the pretrained single image compression model ELIC \cite{he2022elic}. For MS-SSIM evaluation, the MSE-optimized models are fine-tuned for $300$ epochs using the MS-SSIM distortion loss with the initial learning rate as $5e^{-5}$. During inference, we set the number of decoding steps as 8. Our experiments are conducted with NVIDIA V100 GPUs using PyTorch. 

\noindent \textbf{Benchmarks.} We compare our CAMSIC with a variety of traditional and learning-based codecs. These competitive baselines are roughly split into four categories: single image compression (BPG \cite{bpg2014}, ELIC \cite{he2022elic}), video compression (HEVC \cite{sullivan2012overview}, VVC \cite{bross2021overview}), multi-view compression (MV-HEVC \cite{tech2015overview}, LDMIC \cite{zhang2022ldmic}) and stereo image compression (DispSIC \cite{zhai2022disparity}, SASIC \cite{wodlinger2022sasic}, BCSIC \cite{lei2022deep}, ECSIC \cite{wodlinger2024ecsic}). For BPG, we adopt the default x265 encoder without chroma subsampling to independently compress each image. By treating a stereo image pair as a two-frame video sequence, we run HM-18.0 and VTM-23.0 software with \textit{lowdelay P} configuration and YUV444 format to evaluate the coding efficiency of HEVC and VVC. For MV-HEVC, HTM-16.3 software is used to compress stereo images by setting a two-view intra mode. Unfortunately, it only supports YUV420 format, resulting in inferior compression performance. As for learning-based multi-view and stereo codecs except ECSIC, we implement them using the same training procedure. Since the current SOTA method ECSIC requires training on large-width stereo images to achieve the best results, we follow their open source library \cite{wodlinger2024ecsic} to train the ECSIC models. 

\begin{figure}[t]
  \centering
  \subfloat
  {\includegraphics[scale=0.1975]{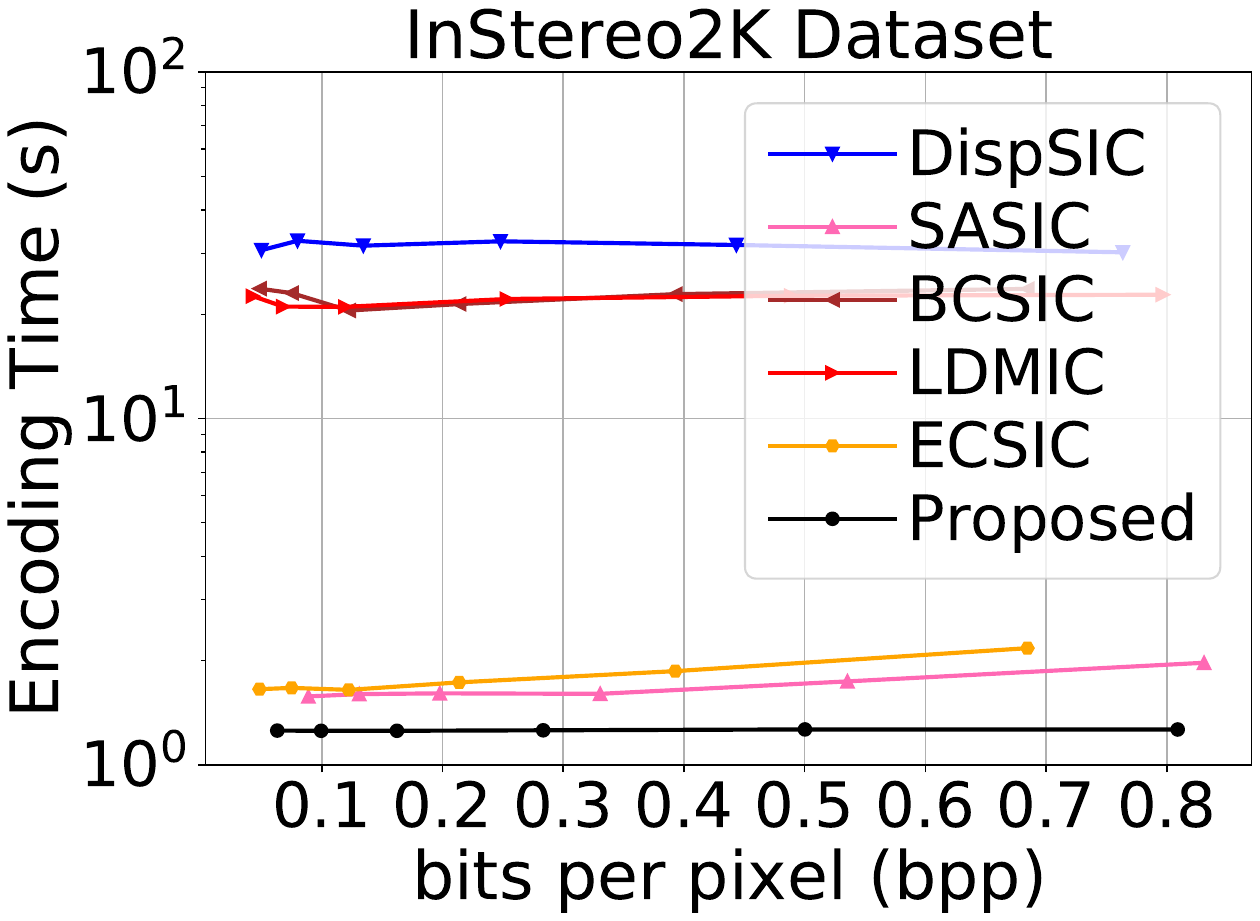}}
  \subfloat
  {\includegraphics[scale=0.1975]{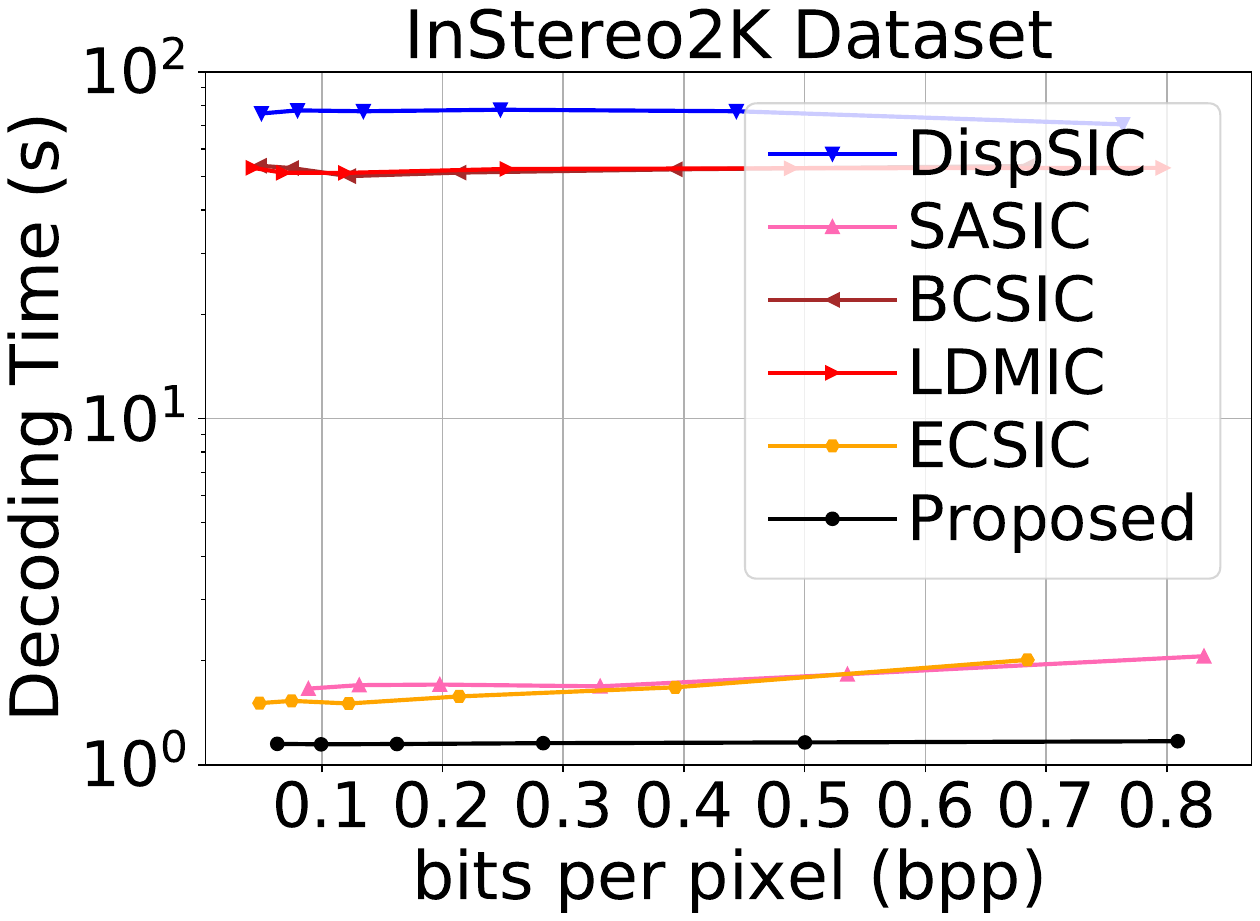}}
  \caption{Complexity of learning-based codecs on the InStereo2K dataset. Both encoding and decoding time are evaluated by an NVIDIA V100 GPU.}
  \label{fig:complexity}
\end{figure}

\subsection{Experimental Results}\label{sec:performance_comparision}

\noindent \textbf{Compression performance.}
Fig.~\ref{fig:rd_performance} shows the RD curves of these methods on Cityscapes and InStereo2K datasets. We report BD-rate, BD-PSNR and BD-MSSSIM scores of each codec relative to BPG in Table~\ref{table:rd_performance}. It is observed that the single image codec ELIC achieves 25.196\% and 29.748\% bitrate saving in terms of PSNR on Cityscapes and InStereo2K datasets. By contrast, our CAMSIC has larger improvements, i.e., 41.961\% and 43.608\%. This demonstrates that introducing a stereo conditional entropy model to capture the correlations between stereo images effectively enables a more accurate estimation of the probability distribution. 

Without manually designed and complex linear or nonlinear transformations, our method outperforms most of these compression baselines, highlighting the importance of a powerful spatial-disparity entropy model. Specifically, a series of unidirectional codecs (HEVC, VVC, MV-HEVC, DispSIC, SASIC) necessitate explicit warping to remove disparity redundancies. Notably, our CAMSIC surpasses VVC, the best unidirectional codec, across all bpp ranges on the InStereo2K and Cityscapes dataset. In addition, bidirectional codecs (BCSIC, ECSIC) adopt bidirectional attention to reduce inter-view redundancies. Remarkably, even with a straightforward image encoder-decoder architecture, our CAMSIC still saves more bits over the SOTA method ECSIC in terms of PSNR and MS-SSIM. Moreover, compared with the latest distributed codec LDMIC that adopts the joint decoding method based on global cross-attention mechanism, our proposed method improves by about 0.419dB and 0.158dB PSNR at the same bpp level on two datasets (Cityscapes and InStereo2K). These results affirm the superiority of our proposed entropy model in exploiting data correlations for enhancing compression performance.


\noindent \textbf{Coding Complexity.}
We compare the computation complexity of six learning-based multi-view and stereo codecs on InStereo2K dataset. The running time on GPU includes the execution latency of entropy coder. As illustrated in Fig.~\ref{fig:complexity}, our CAMSIC method is orders of magnitude faster than other auto-regressive approaches (DispSIC, BCSIC and LDMIC) by leveraging the parallel coding advantages of the MIM technique. Furthermore, by simplifying the stereo image encoding and decoding processes, our approach runs faster than SASIC and ECSIC. These results showcase a better trade-off between performance and speed among contemporary learning-based codecs.

\subsection{Ablation Study} \label{sec:ablation_study}

\noindent \textbf{Different Entropy Models.} 
Table~\ref{table:ablation_study_table} presents the BDBR results of different entropy models on the Cityscapes and InStereo2K datasets. 
Specifically, we replace our entropy model with the CNN-based ECSIC, SASIC and BCSIC entropy models as well as the content-irrelevant encoder-decoder Transformer entropy model. Note that the vanilla MIM-based Transformer entropy model outperforms the ECSIC entropy model, which is attributed to the Transformer's inherent capability to capture long-range dependencies more effectively than CNN architectures. However, simply migrating the vanilla encoder-decoder Transformer architecture and the naive MIM technique to the entropy model leads to inefficient prior information propagation, which makes the vanilla MIM entropy model inferior to the BCSIC entropy model with bidirectional view interaction.
To address this challenge, we introduce a novel content-aware MIM technique to design an advanced decoder-free Transformer entropy model. The improvements demonstrate that our content-aware MIM significantly surpasses the content-irrelevant MIM, which underscores the importance of bidirectional interactions between prior information and estimated tokens in improving compression performance. 

\noindent \textbf{Operations for Token Interaction.} 
In our framework, we enable relative position embedding and introduce identity embeddings to facilitate token interaction. To show the effects of these operations, we undertake a series of ablation experiments. As shown in Table~\ref{table:ablation_study_table}, compared with removing relative position embeddings (variant V5), adding relative position embeddings (variant V6) saves more bitrate, which indicates that providing the position cues for token interaction enables a more accurate estimation of probability distribution. Based on variant V6, we attach two identity embeddings to the input token sequence to form our final version, which further reduces the bitrate consumption. It implies that explicitly distinguishing content-aware tokens and estimated tokens boosts token interaction.



\begin{table}[t]
 \fontsize{8}{10}\selectfont
  \centering
  \begin{tabular}{l|cc}
    \hline
    Variant & Cityscapes & InStereo2K  \\
    \hline
    Ours & 0\% & 0\% \\
    \hline
    (V1) w/ ECSIC entropy model  & 48.89\% & 22.38\% \\
    (V2) w/ SASIC entropy model   & 36.16\% & 19.44\% \\
    (V3) w/ BCSIC entropy model  & 5.36\% & 5.76\% \\
    (V4) w/ Vanilla MIM entropy model  & 36.65\% & 19.48\% \\
    \hline
    (V5) w/o RPE + w/o ID &  3.700  &  1.316  \\
    (V6) w/o ID & 2.739 &  0.501 \\
    \hline
  \end{tabular}
  \caption{Ablation studies for different components. The first row is set as the anchor to measure BD-rate. RPE means relative position embedding. ID indicates identitiy embedding. }
  \label{table:ablation_study_table}
\end{table}

\section{Conclusion}
In this paper, we propose a stereo image compression framework, namely CAMSIC, with a succinct yet potent Transformer entropy model as the core to effectively capture the spatial-disparity correlations. Our CAMSIC consists of a straightforward encoder-decoder architecture to independently transform each image. A novel content-aware masked image modeling (MIM) technique is presented to fully exploit the power of Transformer entropy model. Our content-aware MIM unleashes bidirectional interactions between the prior information and estimated tokens, which allows us to eliminate an extra Transformer decoder. Thus, a decoder-free Transformer entropy model is proposed to improve the efficiency. Experimental results show that our approach achieves the state-of-the-art compression performance with fast encoding and decoding. 

\section{Acknowledgments}
This work was supported by the General Research Fund (Project No. 16209622) from the Hong Kong Research Grants Council.

\bibliography{aaai25}
\appendix

\clearpage

\section{Network Structure}

\begin{figure}[t]
  \centering
  \subfloat[Hyperprior Encoder]
  {\includegraphics[scale=0.88]{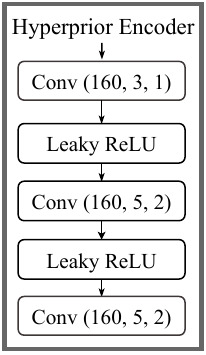}}\hspace{1cm}
  \subfloat[Hyperprior Decoder]
  {\includegraphics[scale=0.88]{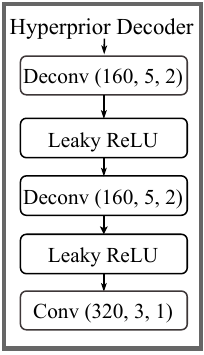}}\\
  \subfloat[Prior Fusion Module]
  {\includegraphics[scale=0.88]{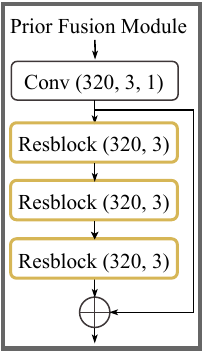}}\hspace{1cm}
  \subfloat[Residual Block]
  {\includegraphics[scale=0.88]{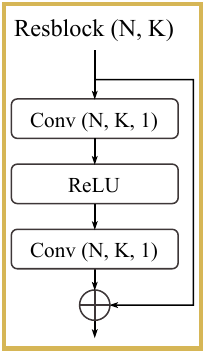}}
  \caption{The network structures of (a) hyperprior encoder, (b) hyperprior decoder, (c) prior fusion module with the details of the residual block shown in (d). “Conv (N, K, S)” represent the convolution operations with the number of output channels as $N$, the kernel size as $K\times K$ and the stride as $S$.}
  \label{fig:em_encoder}
  \vspace{-0.1cm}
\end{figure}

\begin{figure}[t]
  \centering
  \subfloat[Swin Transformer Block]
  {\includegraphics[scale=1]{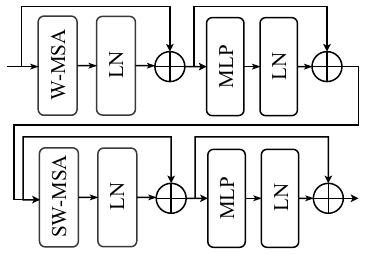}}\\
  \subfloat[Entropy Parameter Module]
  {\includegraphics[scale=1]{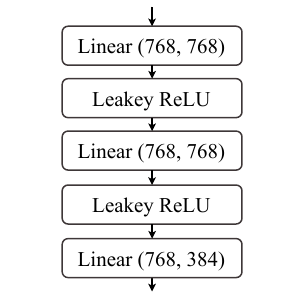}}
  \vspace{0.1cm}
  \caption{The network structure of (a) a Swin Transformer block and (b) entropy parameter module. W-MSA and SW-MSA are multi-head self attention modules with regular and shifted windowing configurations, respectively. LN means layer norm. MLP shorts for multi-layer perception. Linear (input size, output size).}
  \label{fig:transformer}
  \vspace{-0.1cm}
\end{figure}

\textbf{Image Encoder and Decoder.} The network structures of our encoder $E$ and decoder $D$ are the same as ELIC \cite{he2022elic}, which stacks multiple residual bottleneck blocks and simplified attention modules. The number of channels in convolutional layers is set as 320.

\begin{figure}[t]
  \centering
  \includegraphics[scale=0.495]{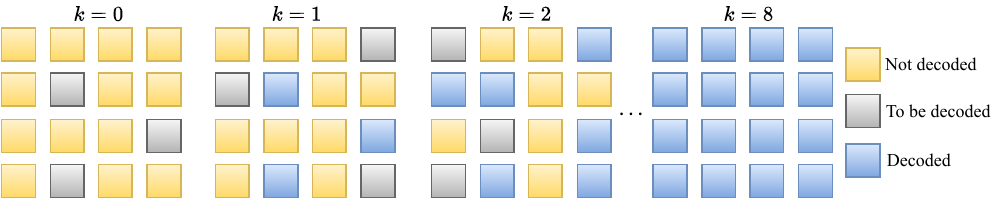}
  \caption{The scheduled parallel decoding of MIM method. During inference, we decode $\hat{\boldsymbol{y}}_v$ in a few iterations. Each iteration recovers a portion of the most certain tokens with the smallest entropy based on the decoded ones.}
  \label{fig:decoding}
  \vspace{-0.2cm}
\end{figure}



\begin{figure*}[t]
  \centering
  \includegraphics[scale=0.52]{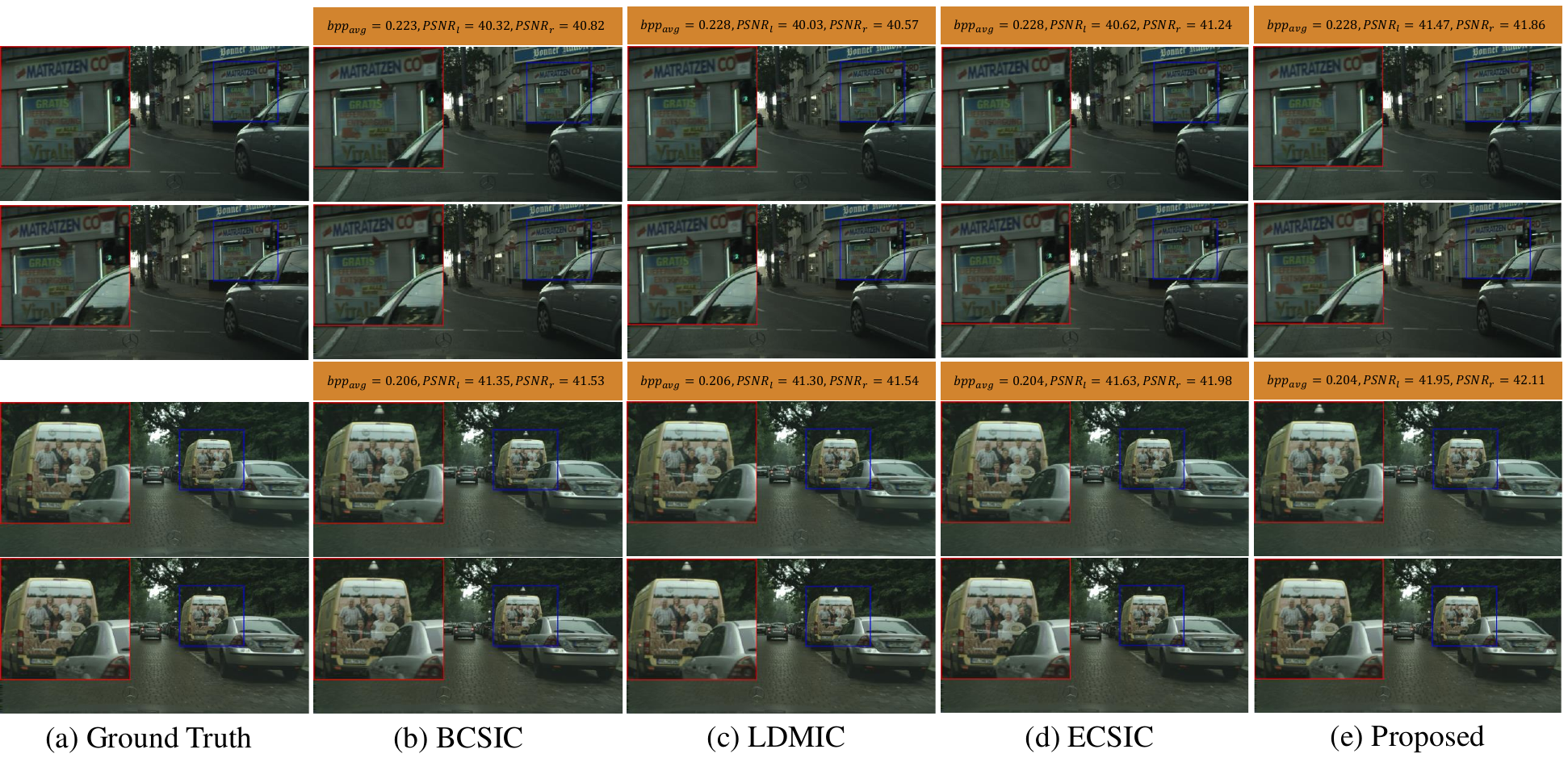}
  \caption{Visualization comparison of our method against other codecs on the Cityscapes dataset.}
  \label{fig:visualization1}
  \vspace{-0.2cm}
\end{figure*}

\begin{figure*}[t]
  \centering
  \includegraphics[scale=0.69]{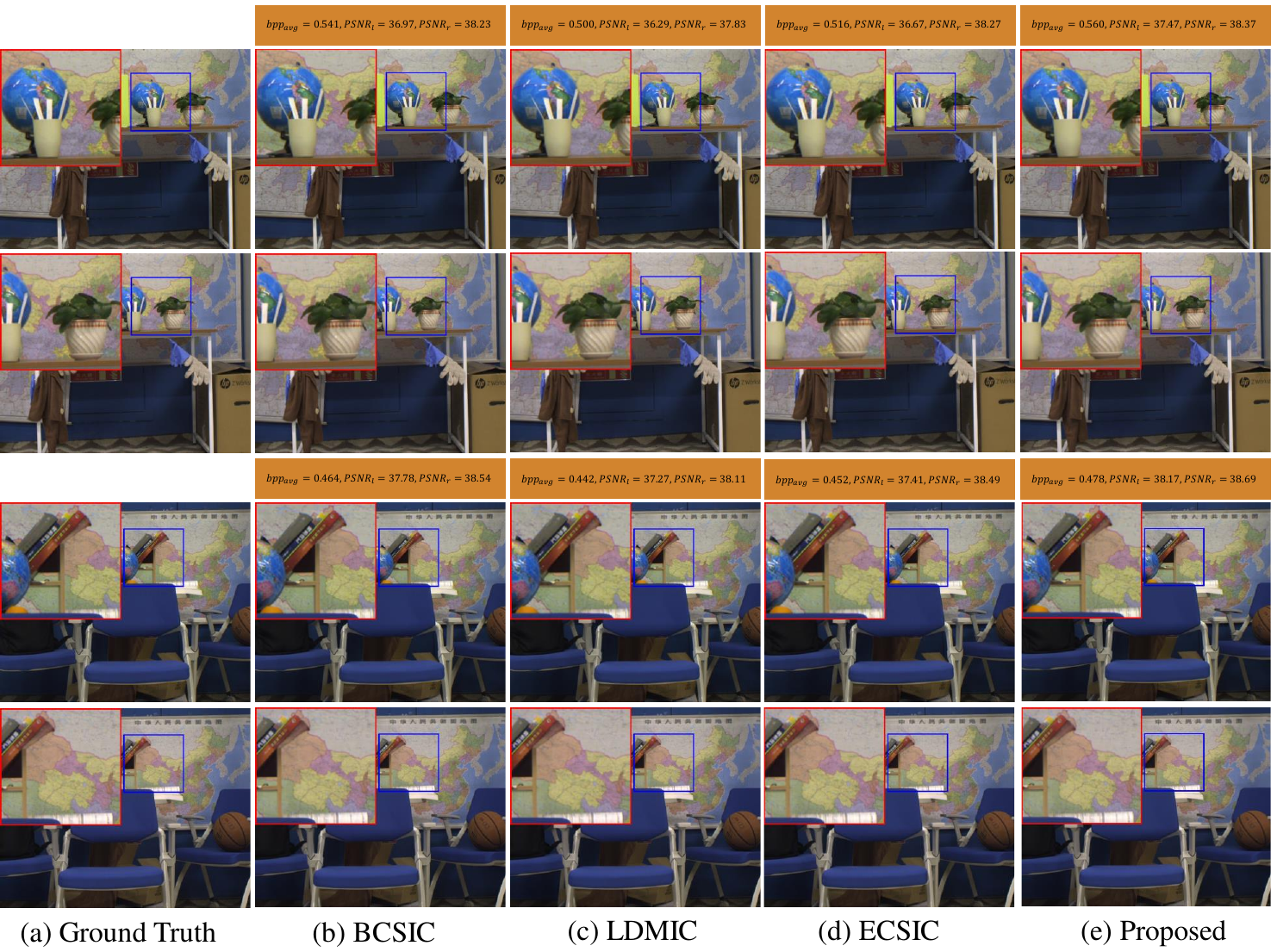}
  \caption{Visualization comparison of our method against other codecs on the InStereo2K dataset.}
  \label{fig:visualization2}
  \vspace{-0.2cm}
\end{figure*}

\noindent \textbf{Prior Generation Network.} Fig.~\ref{fig:em_encoder} shows the details of the hyperprior encoder/decoder and prior fusion module. After generating the hyper prior and disparity prior information, we apply a prior fusion module to produce the content-aware tokens $\boldsymbol{u}_v$.

\noindent \textbf{Decoder-free Transformer.} For the projection module, we use a linear layer to expand the dimensions from 320 to 768. As shown in Fig.~\ref{fig:transformer} (a), we stack four post-norm Swin Transformer blocks \cite{liu2022swin} to form our Swin Transformer encoder. The window size of W-MSA and SW-MSA is 8. The number of attention heads is 12 for every attention layer. We use two layers of MLP with the input size, hidden size, and output size as 768, 3072 and 768, respectively. As for the entropy parameter module, we apply three linear layers with Leakey ReLU activation function shown in Fig.~\ref{fig:transformer} (b). 

\section{Iterative Decoding Process}
We follow the iterative decoding process in \citet{chang2022maskgit} to generate an image during inference. Taking the content-aware masked image modeling (MIM) method as an example, we begin with all the tokens generated by prior information, i.e. $\boldsymbol{c}_v$. For ease of notation, let $\boldsymbol{u}_v^{0}=\boldsymbol{c}_v$. As shown in Fig.~\ref{fig:decoding}, the decoding process for iteration $k$ runs as follows:

\noindent \textbf{Predict:} Given the input tokens $\boldsymbol{u}_v^{k}$ of the decoder-free Transformer at the current iteration, our entropy model predicts the entropy of a token:
\begin{equation}
  \begin{aligned}
    H(\hat{\boldsymbol{y}}_{v}^{i,k})&=- \log_{2} \prod_{j=1}^{d} p(\mu_{v}^{i,j,k}), \\ p(\mu_{v}^{i,j,k})&=  c_{v}^{i,j,k}(\mu_{v}^{i,j,k}+\frac{1}{2})- c_{v}^{i,j,k}(\mu_{v}^{i,j,k}-\frac{1}{2})
  \end{aligned}
\end{equation}
where $d$ is the channel dimension of a token, $i$ is the position of the token, and $c_{v}^{i,j,k}(\cdot)$ is the cumulative function of Gaussian distribution $\mathcal{N}(\mu_{v}^{i,j,k}, {\sigma_{v}^{i,j,k}}^2)$. 

\noindent \textbf{Schedule:} We compute the number of tokens $\left\lfloor \gamma(k)N - \gamma(k-1)N \right\rceil$ to be recovered at each step according to the scheduling function $\gamma(k)=\sin(k\cdot \frac{\pi}{2})$, where $N$ is the input token length. 

\noindent \textbf{Decode:} We decode the selected tokens from bitstream with $(\boldsymbol{\mu}_v^{k}, \boldsymbol{\sigma}_v^{k})$ and generate the input tokens $\boldsymbol{u}_v^{k+1}$ for iteration $k+1$.

\section{Visualizations}

We conduct the visual comparison of our proposed method against BCSIC, LDMIC, and ECSIC. Several examples from Cityscapes and InStereo2K datasets are shown in Fig.~\ref{fig:visualization1} and \ref{fig:visualization2}. From these examples, we see that our approach achieves better rate-distortion performance.

\end{document}